# JADE: a board game to teach software ergonomics.

Stéphanie Jean-Daubias,

Univ Lyon, UCBL, CNRS, INSA Lyon, LIRIS, UMR5205, F-69622 Villeurbanne, France
stephanie.jean-daubias@univ-lyon1.fr

**Abstract.** JADE is an educational game we have imagined, designed, built, and used successfully in various contexts. This board game enables learning and practicing software ergonomics concepts. It is intended for beginners. We use it every year during several hours with our second-year computer science students at Lyon 1 University. In this paper, we present the classical version of the game, as well as the design and evaluation process that we applied. We also present the hybrid version of JADE, which relies on the use of QR codes and videos. We also present its use in our teaching (with about 850 learners for a total duration of 54 hours, which totals more than 2500 student-hours). We then discuss the results obtained and present the considered evolutions.

**Keywords:** ergonomics, assessment criteria, cases study, pedagogical game, board game, hybrid game, QR code.

## 1 Introduction

Being both a software ergonomics teacher and a researcher in educational software for more than twenty years, we are always looking for ways to innovate and to improve our teaching. One of them is JADE: a board game we have designed to teach software ergonomics.

**Motivation.** The teaching of HMI (Human Machine Interaction), and in particular software usability, is increasingly present in computer science training: at university, in engineering schools, and also in other post-graduate training. These courses are important, as a complement to the teaching of programming, so that computer scientists will not only be able to develop useful programs, but also usable ones, adapted to their users. Learning the basic concepts of software ergonomics generally combines theoretical teaching and practical application, particularly through case studies. Despite these practical situations, some concepts are sometimes quickly forgotten by the students once they have passed the final exam, or even before...

In order to anchor the concepts of ergonomics more durably in the minds of our second-year students of the computer science degree at the University of Lyon 1, we have sought to integrate a playful situation into our teaching. The positive impacts of using games in an educational context are multiple and recognized: motivation and active participation of learners, interaction and cooperation between learners, consolidation of knowledge, and encouragement of problem solving in particular [1].

**Approach.** In contrast to the current trend of digital serious games, we have chosen to create a "physical" board game, unplugged, which, in addition to the advantage of encouraging human interaction between students, takes students away from their computers for a while.

**Research Context.** *Game* is defined by Caillois [2] as a free, separate, uncertain, unproductive, regulated and fictitious activity. *Serious games* are still sometimes wrongly considered to be an oxymoron. Although their definition can vary from one author to another [3] [4] [5], they can be broadly defined as follows: a serious game is a device, digital or not, whose initial intention is to combine serious aspects with playful elements [6]. *Learning games* are a subset of serious games, those dedicated to learning [7]. The corresponding scientific field, Game-Based Learning, has been developing increasingly since the 2000s. If the vast majority of the games developed are digital (which is particularly true for IT-related courses, and in higher education), in this article, however, we focus on *physical educational games*.

Non-digital educational games are more and more used in education, even if they are less widely used than digital educational games. Among physical educational games, *card games* are frequently chosen, probably because they are familiar to learners and relatively easy to make, handle, store, and transport. The rules may be simple, so that the game is easy for players to get familiar with. When the learning situation is more complex, physical educational game designers tend to prefer *board games*. These games offer a richer environment, even if they are harder to design, manufacture, store, transport, and handle. Their rules are often much more complex, and therefore it takes longer, and it is more difficult for players to get familiar with them.

The games used for education range from existing non-educationally intended games used as is, to innovative games created from scratch. Some teachers use indeed existing games for their pedagogical features, directly in their teaching practices (for example Spot-it [8] or Chromino [9] for primary school mathematics, Scrabble [10] for language teaching). Others adapt existing games to their subject by changing the content (text and pictures) but keep the game rules (for example Timeline [11] of the History of computing [12], 7 families [13] of famous computer scientists [14]). Finally, other teachers design their own games so that they are perfectly suited to their needs. These new games are usually based on existing, often classic, board games (as Goose game [15], which leads players step by step towards a goal, or Trivial Pursuit [16], which allows players to demonstrate their mastery, even partial mastery, of different subjects).

Physical educational games are mainly found in elementary education. The older the pupils are, the fewer physical games there are. In higher and continuing education such games are less common, and even less if we focus on information technology training. Indeed, when IT training courses opt for learning games, the vast majority choose digital approaches. So, it's not surprising that we haven't found neither game that met our needs, nor even any that was close to our objective.

In the following of the paper, we present the classical version and the hybrid version of the game we conceived in this context, as well as the design and evaluation process that we adopted, and the numerous uses that have been made. We end this paper with discussions, conclusions, and perspectives.

## 2  Presentation of the Game

### 2.1 Audience and Educational Objectives

**Context.** We teach HMI and interface ergonomics to students in the second year of a computer science degree at the University of Lyon 1. Our teaching unit lasts 10 weeks. It consists of 30 hours divided into 15 hours of lectures and 15 hours of practical work. The lectures include from 70 to 150 students depending on the semester, the practical work involves 40 or 20 students depending on the session. For the practical work, several groups of students work in parallel. Each week, students attend 1.5 hours of lectures immediately followed by 1.5 hours of practical work on the same subject. The examination takes place in the eleventh week. The teacher is a university full professor teaching this subject for more than twenty years, she is assisted by 2 to 3 teaching assistants who change regularly.

**Audience.** Students who attend this course have a scientific baccalaureate/science A-level and are currently studying for a degree in computer science. Most of them have already completed the first year of this degree and are therefore in their second year of a computer science degree where they learn programming. Nevertheless, these students have no prior knowledge of software ergonomics.

**Objective of the Teaching Unit.** The teaching unit aims at teaching HMI. It mainly deals with the notions of design, mock-up, evaluation, user testing, and ergonomic analysis.

**Research Questions.** In this project, the author of this article has a dual role: she is both the client, as a teacher in software ergonomics, and the designer of the device, as a researcher in educational software. The issue is therefore also twofold: (1) as a teacher, our problem can be formulated as follows: How can we facilitate both the learning of ergonomics and the motivation of students; (2) as a researcher, our problem can be formulated in the following way: How can we adopt a playful approach at university while encouraging direct interpersonal interactions?
   In relation to these issues, the research questions that guided our investigation were:
   Q1: What types of games are suitable for practice and argumentation?
   Q2: What role can games play in a university course?
   Q3: What types of game are suitable for computer science students?

**Objective of the Game.** Before starting to create a game, we carefully identified the intended objectives [17]. They are twofold. On the one hand, it is about encouraging the learning of theoretical concepts taught in lectures and putting them into practice, in the idea of "concretizing the abstract". On the other hand, it is a question of encouraging the active participation of students, their interactions and feedback. We intend to improve their motivation and their commitment, in a benevolent, pleasant, and relaxed context. Research has shown that gamification of instruction is suitable for addressing these objectives [18].

**JADE Game.** The game that we designed, built and that we now use in our classes is named JADE (acronym for "Jeu d'Apprentissage De l'Ergonomie", which means "game to learn ergonomics" in French). JADE is a board game where teams compete to become the best software ergonomics coach team. To do this, players apply their theoretical knowledge to concrete cases. They identify usability problems, explain them by arguing with an opposite team and propose appropriate solutions to correct them. The game is mainly aimed at students entering the field of ergonomics.

### 2.2 Composition of the JADE Game

JADE [19] is a physical board game made up of a set of elements: the game board itself, several software boards, the sheet of pictograms legend, pawns for each teams, one dice, a pencil, and scores sheets. The game is stored in a A3-size pocket (42 × 30 cm), easy to handle in classrooms and to distribute to students.

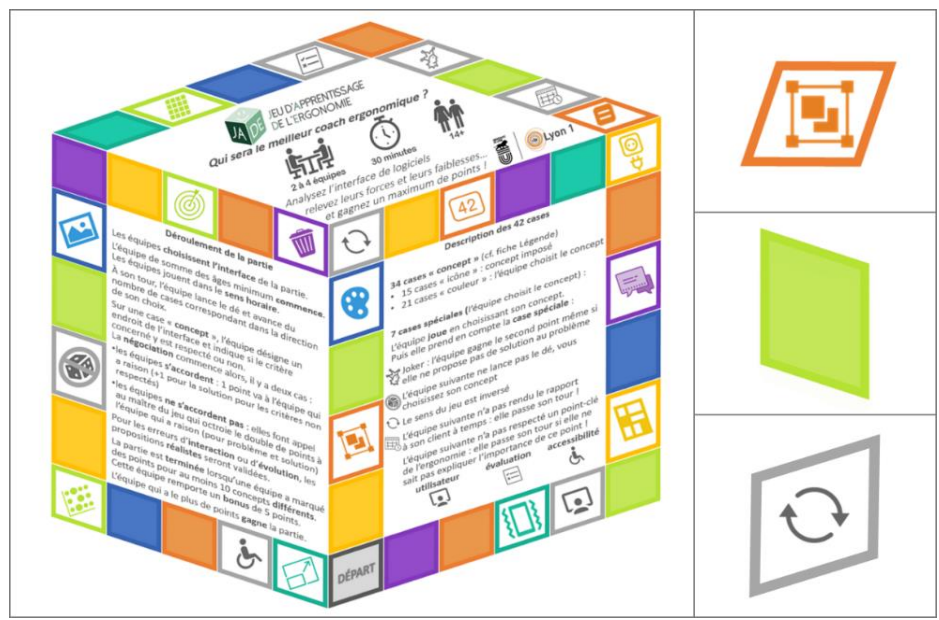

**Fig. 1.** Multicolored board game of JADE (left) and zoom on a concept box, a color box, and a gamification box (from top to bottom right).

**Game Board.** The game board (cf. figure 1) is similar to a trivial pursuit board, it reflects the progression of the teams in the game: the concepts to be studied and the play events. The board has a cube shape (that evokes a Rubik's cube), it comprises 42 boxes (a nod to *The answer to the ultimate question of life, the universe, and everything* [20]) of different types.

**Concept Boxes** contain a pictogram corresponding to the ergonomic concept to be studied (cf. example in the center in the right part of figure 1). For each concept, we have chosen a pictogram that represents it, in the color symbolizing this course (see below). The pictogram is used in the course slides and then in the game for students to become familiar with them.

**Color Boxes** (including the starting box) allow a team to choose between all the concepts of this color, that is to say between this family of concepts listed in the legend (cf. example in the middle right of figure 1). This choice could seem to be an easy way for the students, but it is in fact a way of encouraging them to work better on all the concepts of the course: indeed, to choose a concept, players should first master them all enough to choose the one which will be the more favorable for them in the game!

**Special Boxes** are gamification boxes represented with gray pictograms on the boards (from 0 to 7 depending on the version of the board). When it falls on a special box, the team plays normally by choosing its concept, then it takes into account the special box:
1. *Joker*: the team always win the second point (cf. gameplay in section 2.4).
2. *No dice*: the next opposite team does not roll the dice: the current team chooses the concept they will play with.
3. *Reverse direction*: the direction of the game is reversed; the current team plays a second time (cf. bottom right of figure 1).
4. *Late*: the next opposite team has not returned its report in time: it loses its turn.
5. *Key-points*: the opposite team has not respected a key point of ergonomics: they pass their turn unless they can explain the importance of this point (these boxes are an opportunity to recall fundamental concepts in ergonomics: *role of user*, *assessment*, *accessibility*).

There are six versions of the game board to take into account the progress of students in the six steps of our ergonomics course (general theories for ergonomics, guidelines for GUI elements, general ergonomic criteria, ergonomic criteria for the World Wide Web, mobile and tactile ergonomic criteria, ergonomic criteria for disability), plus a seventh combining the first six (cf. figure 2).

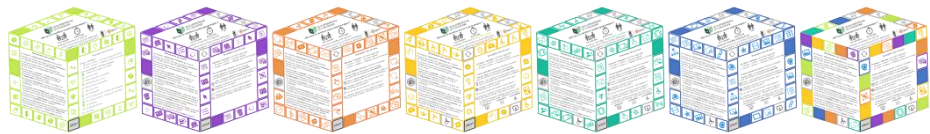

**Fig. 2.** Overview of the 7 versions of the board game of JADE.

**Lime Green Board Game: General Theories for Ergonomics**. The first week the students use the lime green board game that contains only the concepts of the first ergonomics course which presents the general theories for ergonomics. The content of this course is divided into twelve elementary units suited to the JADE game: *Miller's magical number*, *Hick's law*, *two-second rule*, *three-click rule*, *baby bird syndrome*, *affordance*, *gestalt psychology-law of proximity*, *gestalt psychology-law of similarity*,

*rules about colors*, *reading direction*, *Fitts's law*, and *rules about text*. Each concept is present as least twice on the board, most important concepts or those which are difficult for students are present three times. There are no gamification boxes in this version of the game to allow students to focus on the basic rules of the game during the first session.

**Purple Board Game: Guidelines for GUI Elements**. The second version of the JADE board game contains only the concepts of our second ergonomics course which deals with guidelines for GUI (Graphical User Interface) elements. We have grouped the most common components into eleven categories: *windows*, *dialog boxes*, *icons*, *menus*, *pointers*, *buttons*, *lists*, *groups*, *input areas*, *wheels / selectors*, *progression bars*, the twelfth concept is dedicated to *feedback*. This limitation that the number of items per category should be a multiple of six was inherited from an early version of the game and is now deprecated, but we haven't remade all the items in the game since. In addition, this board game introduces four gamification boxes (see 1 to 4 in the above list) to enrich the game. This gradual introduction of gamification boxes allows students to progressively become familiar with their functioning.

**Orange Board Game**: **General Ergonomic Criteria**. The third version of the JADE board game is devoted to general ergonomic criteria (i.e., not specific to a particular type of application). We use Bastien and Scapin's criteria [21] in our course because we find them very relevant and convenient from a pedagogical point of view. In the JADE game, we don't only use Bastien and Scapin's eight main criteria (in capital letters in the following), for pedagogical reason we prefer to use most of the sub-criteria proposed by the authors. We have therefore chosen 18 criteria or sub-criteria: GUIDANCE (*prompting*, *grouping/distinction of items*, *immediate feedback*, *legibility*), WORKLOAD (*concision*, *minimal actions*, *information density*), EXPLICIT CONTROL (*explicit user action*, *user control*), ADAPTABILITY (*flexibility*, *user experience*), ERROR MANAGEMENT (*error protection*, *quality of error messages*, *error correction*), CONSISTENCY, SIGNIFICANCE OF CODES, and COMPATIBILITY (in which we distinguish in the game *adaptation to users* from *compatibility between applications*). From this version of the game, the boards include the seven gamification boxes (see 1 to 7 in the above list) as students are now familiar with how they work.

**Yellow Board Game**: **Ergonomic Criteria for the Web**. The next version of the JADE board game is centered on ergonomic criteria specific to the Web. For this part of our course, we have chosen to use Amélie Boucher's criteria [22]: they are not rigorously scientifically validated (unlike the previous criteria), but they are convenient in an educational context. We therefore use in JADE these twelve criteria: *architecture*, *visual organization*, *consistency*, *conventions*, *information*, *understanding*, *guidance/assistance*, *error management*, *speed*, *freedom*, *accessibility,* and *internet user satisfaction*. Most of them are consistent with Bastien and Scapin's criteria, but we focus in our course on the specificity of the Web for each of these criteria.

**Emerald Board Game**: **Mobile and Tactile Ergonomic Criteria**. This version of the JADE board game is centered on ergonomic criteria specific to mobile devices. As we did not find in the literature any specific criteria suitable for our teaching, we defined our own heuristics [23]. As the previous ones, those criteria are not scientifically validated, but they are suited to our needs in our course and in the game: *adapt the display to the screen*, *facilitate the tactile click*, *limit the quantity of information*, *avoid*

*use of the keyboard*, *use adapted components*, *allow the recovery of errors*, *exploit the specificities of mobile*, *ensure continuity between devices*, *speed up navigation*, *think about accessibility*, *provide a degraded mode*, *inform about authorizations*. The color of this board comes from the name of our ergonomic heuristics for mobile and tactile. Indeed, they are called EMERAUDE, the French word for emerald, an acronym whose English equivalent could be "Ergonomics for Mobile devices: Establishing Rules to increAse the Usability of Environments".

**Blue Board Game**: **Ergonomic Criteria for Disability**. The last game board is dedicated to ergonomic criteria related to disability and digital accessibility. In our teaching we use French guidelines close to the W3C recommendation WCAG 2.0 (Web Content Accessibility Guidelines) [24]: the RGAA (for *Référentiel Général d'Amélioration de l'Accessibilité*: general repository for improving accessibility) published by the French government [25]. The criteria are as follows: *images*, *frames*, *colors*, *multimedia*, *tables*, *links*, *scripts*, *mandatory elements*, *structuring of information*, *presentation of information*, *forms*, *navigation*, and *consultation*. The color of this board comes from the color usually associated with disability.

**Multicolored Game Board**: **Combination of all the Concepts of the Different Courses**. A last board game combines the colors and pictograms of all the previous boards. It allows to approach in a same game session, all the concepts of the different courses and to combine them. To allow a good representation of all the concepts of the courses, the most important concepts, the most difficult or usually misunderstood by the students are represented on certain boxes, the other boxes are color boxes which leaves the choice of the concept to the players. Our seven gamification boxes are also present on this board.

**Software Boards.** These are double-sided sheets in A3 format (42×30 cm) laminated. Each side presents a software, website, or mobile application. The pictograms at the top left indicate which versions of the boards it is compatible with (cf. figure 3).

A software is represented by its main screens, the sequence between the different screens is symbolized by arrows, sometimes supplemented by labels (for example to indicate the time between the display of two screens or the conditions necessary to pass from one screen to another). These indications help players to understand the functioning of the software and its interactions with the user. There is a substantial number of screens and indications in a software board to give as much material as possible to the teams to study the software from the point of view of the different courses on ergonomics.

There are currently 15 software boards to cover the needs of the different game boards. It is easy and quite fast to add such new boards, to enrich the game and to offer students a wide variety and diversity of content.

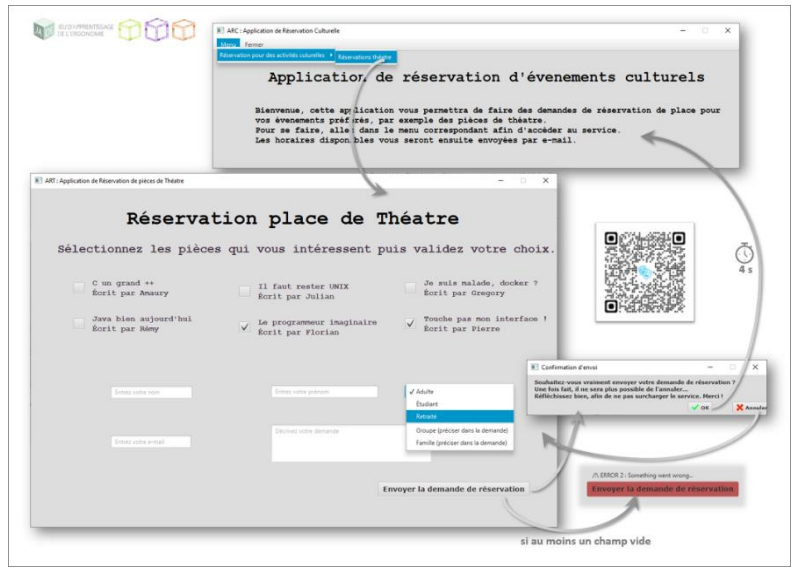

**Fig. 3.** Example of software board: a "toy" software suitable for lime, purple and orange JADE.

**Legend: Synthesis of the Courses.** The legend of pictograms sheet (cf. figure 4) lists all the ergonomic concepts used in the different versions of the game. They correspond to the software ergonomics concepts taught in all the courses. Teams use them to know which concept they should study or to choose their concept, while making the link with their courses.

**Fig. 4.** JADE legend of pictograms.

**Scores Sheets.** For each version of the board game, we made the corresponding score sheet: these A5 sized sheets list all the concepts or categories of concepts dealt with in the board game used (cf. figure 5). The students first write the composition of the teams on it. Then teams tick the corresponding boxes when they score points. So, we have to print and bring the right number of score sheets of the right "color" for each session. We distribute them to the players at the beginning of the session.

**Fig. 5.** Examples of scores sheets for lime JADE (left) and multicolor JADE (right).

**Rules of the Game.** The rules of the game are written in the center of the game board, but they are also available on a web page that the students have to read before the first practical session with the JADE game. In addition, several videos present the game: first, some students have made a video presenting the game and second, we have created a stop motion video to explain the functioning and the rules of the game.

**"JADE recipe" Sheet.** A last sheet presents the organization of the session. It is organized in four sections: before, during, after and "what's next?". "Before", "during" and "after" explain how to position the tables of the room to play, how to tidy up the equipment after the game, recommends to behave quietly to respect the classical work in the adjacent rooms, etc. "What's next?" is the debriefing section (cf. section 5).

### 2.3 Actors and Interactions

There are two kinds of actors in the JADE game: players and game master.

The **players** are the students. They play in teams of two to allow two kinds of interaction (cf. figure 6). The two members of a team interact together to define their answer, in a cooperative way. Then, they interact with the members of the opposite teams to assert their point of view, in a competitive way.

The **game master** is the teacher. He or she follows the progress of the game and can intervene to take position between the teams if they cannot come to an agreement (see below). In addition to the first two types of interaction, there is thus a third type of interaction: the interaction between the learners and the teacher who can intervene to guide the students during the game or intervene as game master.

## 2.4 Gameplay

The aim of the game for players is to become the best ergonomics coaches by competing against one or more opposite teams of two ergonomics apprentices. For this, teams will have to demonstrate their skills in the different fields of software ergonomics through concrete cases of software evaluation.

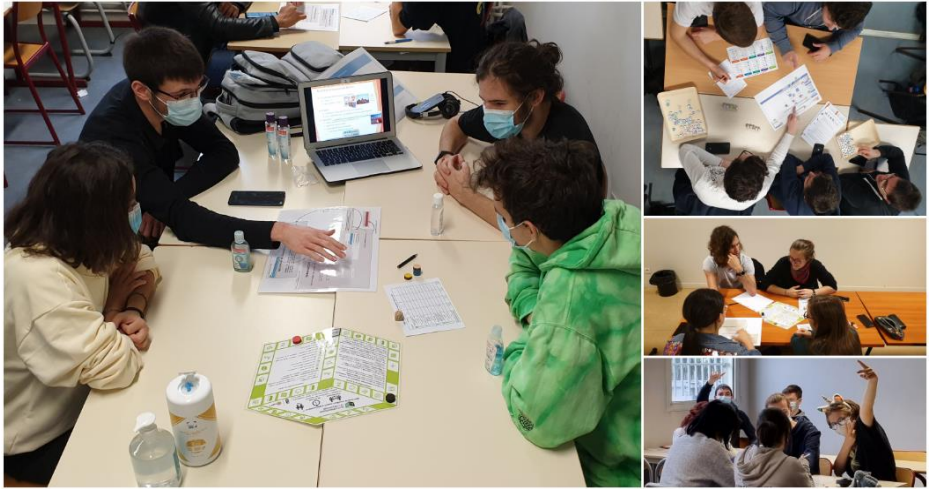

**Fig. 6.** Uses of different versions of JADE in practical courses by students: V1.1 (top right), V2.0 (middle right) and V2.2 (left and bottom right).

**Choice of the Board Game.** The participants start by choosing the software board to study, if it is not imposed by the game master.

**Teams and Turns.** Each team is represented by a pawn on the game board, beginning from the starting box. The players decide among themselves which team goes first, then each team plays in turn. JADE also has a playful way to decide which team plays first. Someone rolls the dice, depending on the draw, the team that starts is the one: (1) who wins at the paper scissors stone game, (2) of the first in alphabetical order of first names, (3) of the last in alphabetical order of first names, (4) of the first per day of birth (+ month if tied), (5) of the last per month of birth (+ day if tied), (6) of the one who makes the greatest number with the dice. This technique allows players to get to know each other and to start the game in a relaxed atmosphere.

**Movements on the Board Game.** The first team to play rolls the dice, decides in which direction they move their pawn, and move it forward by the number of boxes indicated on the dice. The draw of the dice combined with the choice of direction on the game board gives the concept of ergonomics to be studied. The direction of movement in the game is deliberately free: thus, depending on the configuration of the game, the players have the choice between 2, 3, or even 4 different boxes and therefore different concepts. What could be seen as a help to the teams, is in fact an incentive to work harder. Indeed, to choose between the 2 to 4 possibilities available to them, students must have sufficient knowledge on the corresponding concepts, which obliges them to study them carefully if they are not mastered beforehand. In addition, the players have sometimes the opportunity to choose the concept they will have to study (among all the concept of the color, that is to say up to 18 concepts) if they land on a "color" box, which further increases the choice, but also the concepts to be studied.

**Study of a Concept.** Once the concept has been identified, the team that plays must find an element of the interface on the software board that does not respect the given concept, justify their choice to the opposite team and propose a solution to the identified problem. If the team does not find any element in the software that does not respect the concept to be studied, it can describe an element that respects the concept. In this case, the team cannot earn 2 points, but only the first one.

**Scoring.** At each turn, a team can win 2 points (or even 3 in the multicolored version of JADE):
1. The first point is given if finding a problem in the software that corresponds to the concept to study (or a correct application of the concept) and explaining it properly;
2. A second point is given for proposing a relevant solution to the identified problem;
3. A third point may be given, depending on the instructions given by the game master, either to find and explain another concept linked to the studied concept (in order to show the relationships between concepts taken from the different courses) or to find a problem concerning another concept created by the proposed solution (in order to encourage students to take a step back during ergonomics assessment by showing the wider impact of local improvement).

**Argumentation and Refereeing.** It plays an important role in the game, as points are only awarded if the opposing team is convinced by the players' arguments. In case of disagreement between the teams, they can refer to the game master. The game master acts as a referee, has the power to arbitrate between the teams and to distribute the points as he/she thinks most appropriate. When the game master is called, the points are doubled and may be awarded to any of the teams (the playing team or the others). Let's take an example with a game between teams A and B: team A is playing, and the two teams cannot agree on the validity of team A's arguments. The two teams call on the game master and present their arguments one after the other. As team B is completely right in that case, the game master gives 2 points to team B for the problem, plus 2 points for the solution. But if the A team had been right, it would have won the 4 points. So, the players had better agree among themselves instead of appealing to the game master. On the one hand this game rule encourages students to demonstrate argumentative skills and on the other hand helps the game master to manage his/her time in the classroom. Indeed, the teacher is sometimes alone to manage up to 40 students, the autonomy of the latter is necessary in this situation and this game rule is a pedagogically relevant solution to manage the teacher's low availability.

**Replayability.** JADE is usually used during sessions of 1.5 hours, every two weeks. At each session, the game board changes, but some software boards remain the same. In order to prevent students from getting bored, we have created several different software boards for each game board. This is particularly important in the case of repeater students: it is important that there are enough different software boards so that they do not have the impression that they have already done the game. In addition, the diversity of the squares on the game boards already ensures a good replayability of JADE.

## 3 Design and Evaluation Process of JADE

### 3.1 Game Design

JADE (cf. figure 7) is a pedagogical game imagined by a university teacher, created in interaction with two of her colleagues, and for some aspects of the game design, in collaboration with students.

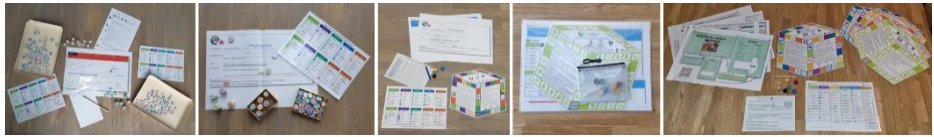

**Fig. 7.** Evolution of sets of the JADE game (V1.0, V1.1, V2.1, V2.2 in the pocket, and V3).

The design started from a desire to get students to work on ergonomic concepts in a practical and fun way. The initial idea of the game came to us while playing TatOuvu ? [26] with our children, a French thinking and observation game similar to Lynx [27]. The goal of these games is to spot the image depicted on the card in play on the game board. The first player to place his token on the corresponding image wins the round.

We thought of using the game boards to show the software that players would have to evaluate in terms of ergonomics. We wanted to replace the small objects to be recognised in Lynx by the ergonomic concepts to be identified in the software. So we listed these concepts and associated them with icons so that they could be represented in the game. We then looked at how the concepts would be assigned to the players during the game. To do this, we imagined several solutions based on different board games that we knew and that we felt were suited to our project and our needs (including dice games, Trivial Pursuit, Monopoly, Snakes and Ladders). The very first version of JADE, closer to Lynx than the current version, was based on a large number of dice associated with tokens representing the concepts to be placed on the software; the current version, deliberately simpler in terms of fabrication and manipulation, is based on a game board on which the players advance the pawns according to the draw of a single die.

Finally, the JADE game is very far from our initial idea because the design process, alternating design, production and real-life testing phases, has led us to significantly make the game evolve towards the current solution, which is much more suited to our needs and those of our users.

The physical game itself is homemade, it was made by the author at home with the help of her family with easily available materials (A4 and A3 format paper, wooden dice and tokens, laminator sheet, A3 format pockets, tote bags) and equipment (laser printer, laminator, cutting machine, cutting ruler, cutter, glue, permanent markers).

The design took place between 2017 (date of original idea) and 2022 (date of the current hybrid version, cf. horizontal arrow in figure 8). We adopted an iterative design method based on a spiral model [28]. Figure 8 shows the design steps and the corresponding versions of JADE (cf. figure 6). In each phase of the cycle, we determined our objectives, evaluated the alternatives and made choices, developed and tested the chosen solution. The very first versions were tested in an informal context with the author's teenagers, and then in a practical session with volunteer students. The following versions were tested in real-life practical works with our students of computer science degree, first with a few students, then more broadly (cf. vertical arrow in figure 8). Since fall 2021 semester, JADE is fully integrated in the teaching, as normal practical works. It is used 2 semesters per year, 4 times per semester with over 150 students per year.

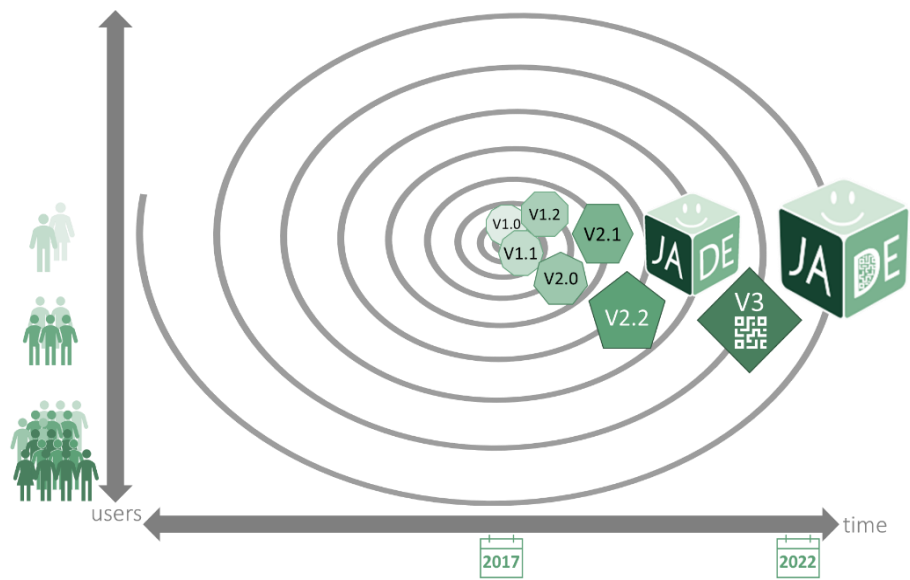

**Fig. 8.** Design cycle of JADE.

Several versions of JADE have been developed successively (cf. figure 6 and figure 7), each new version taking into account the feedback from the tests of the previous one. The first version of the game (JADE V1) was more complex than the current version and required too much material, which was very time consuming to make, but also required a lot of handling time to set up the game and especially to put it away by checking that the games were complete. We made several adjustments to this version of the game (V1.0, visible on picture 1 in figure 7, to V1.1, visible on picture 2, and V1.2) before letting it down totally in favor of JADE version 2, which takes into account the lessons learned from our first experiences. JADE V2 (cf. pictures 3 and 4 of figure 7) is simpler, both in terms of material and game rules. This is the version that was presented in section 2. JADE V3 (cf. last picture of figure 7) is the hybrid version that we will describe in section 4.

### 3.1 Assessment

At the end of most uses, students responded to a brief satisfaction questionnaire. This questionnaire (cf. fig 9) aims at encouraging students to take a step back from the session and find out how they feel about the game, using closed questions, and to gather their comments and suggestions on the game using open questions.

> **Did you find JADE easy to use?** (No, not at all / Not really / Yes, quite / Yes, totally)
> **Did you find the game rules easy to understand?** (No, not at all / Not really / Yes, quite / Yes, totally)
> **JADE allowed you to practice ergonomics:** (No, not at all / Not really / Yes, quite / Yes, totally)
> **JADE enabled you to work on argumentation:** (No, not at all / Not really / Yes, quite / Yes, totally)
> **JADE enabled you to apply the course:** (No, not at all / Not really / Yes, quite / Yes, totally)
> **Comments** (3 blank lines to write a free answer)
> **Suggestions** (3 blank lines to write a free answer)

**Fig. 9.** Questions from the questionnaire to be filled by the players after a session with JADE.

We received approximately 220 responses to these paper questionnaires (1 questionnaire per table of 4 players), which we studied carefully after each session.

These questionnaires show that the majority of players find JADE easy to use, find the rules of the game fairly understandable, are aware of having worked well on ergonomics and have put ergonomics into practice. However, players are less aware of having worked on argumentation. The comments are mainly compliments and encouragement to continue using JADE. Suggestions sometimes include ideas for enriching the game or making it more complex.

More recently, we have also added questions about JADE to the learning experiences questionnaire conducted by our university concerning our HCI course (cf. figure 10 for December 2022 questionnaire.

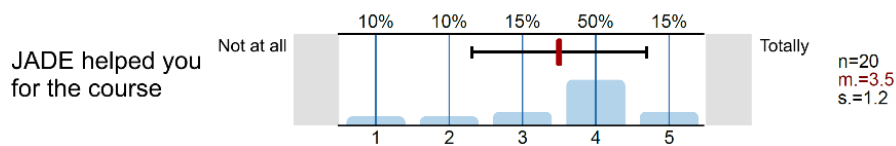

**Fig. 10.** Question about JADE in the December 2022 learning experiences questionnaire.

The answers to these questionnaires, as well as informal live discussions with students, helped us to identify areas for improvement, particularly with regard to the rules of the game. This feedback allowed us to adapt the rules of the game to the students' play skills. Indeed, some students have a very good gaming culture, while others never play board games. In fact, despite the significant momentum of board games over the past two decades, we have observed that some of our students does not have a gaming culture. To adapt the game to this, we simplified its rules for the first boards used during the first sessions with JADE, and we gradually enriched the rules with each new board. Another problem we progressively identified is that players experienced a lack of information on certain software and a lack of information on their interactions, despite the indications added on the software boards. To solve this problem, we decided to enhance JADE, this evolution is presented in the next section.

Another source of feedback came from a demonstration we did at the French HMI conference [29]. The feedback from colleagues who are teachers of software ergonomics was extremely positive and encouraging.

However, we did not conduct a comparative evaluation [30] for ethical and practical reasons: as this game is part of a graded course which has a limited number of teaching hours and which is required for the students' computer science degree, we did not want to create an inequality between students by teaching with or without JADE and then observe the impact of the game on the students' marks in the exam. Otherwise, we studied the students' feedback on the game and on the course, both immediately after the game sessions and at the end of the course (after the final exam).

As reported before, the feedback is very positive: the game creates a positive atmosphere in the group and good motivation, students work more during the sessions with JADE than students do during the classical classes. Moreover, the students express themselves more during the sessions with JADE and work a lot on argumentation. Furthermore, the game facilitates interactions between students who do not know each other beforehand. This was particularly noticeable when students returned to the classroom after the lockdown periods during the Covid-19 health crisis. We also observed that the game, through the interactions it provokes between students, in particular when the teams are heterogeneous, encourages students who are less motivated initially to work more. JADE also allows weak students to improve their skills more than they would have done in a classic practical session. This is partly due to the exchanges between students within a team and between teams: the game forces students to express themselves, even when they are not at ease, and beyond the training that this provokes, the game also makes it possible to correct students' misconceptions.

In order to assess the relevance of our game in relation to our initial objective, we also applied Lanarès, Laperrouza and Sylvestre's FAIR model [32] to the 3 main teaching strategies that we use in our course (lectures, traditional practical sessions and practical sessions with JADE).

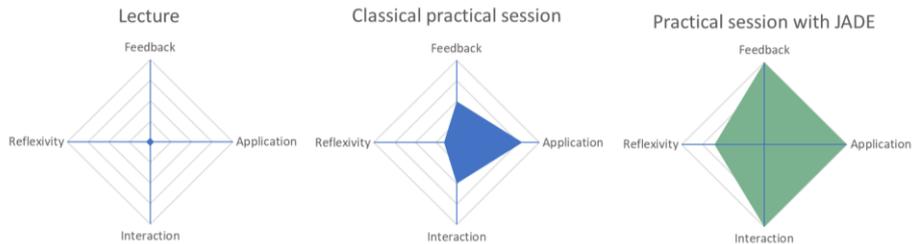

**Fig. 11.** FAIR model applied to lecture, traditional practical session, and session with JADE.

The FAIR model makes it possible to self-assess (and compare) the consideration given to four fundamental principles of learning (clockwise from top in figure 11): behaviorism [33] through the feedback dimension, cognitivism [34] through the application dimension, socio-constructivism [35] through the interaction dimension and constructivism [36] through the reflexivity dimension.

Lectures do not involve any of the 4 dimensions. Classical practical sessions strongly involve the application dimension (these sessions are practical applications of the course), to a lesser extent they involve the feedback and interaction dimensions (via the teacher's interventions during the session or afterwards, if the practical session is graded), and involve the reflexivity dimension only to a very limited extend (in cases where students take the time to reflect on the feedback they have received from the teacher for a graded practical session). The practical session with JADE, meanwhile, involves all 4 dimensions to a considerable extent: it's a course *application* session, extremely *interactive* with significant *feedback* (between learners and with the teacher), it also includes a measure of *reflexivity* (notably via the final debriefing).

## 4 The Hybrid Version of JADE

The use of JADE V2 has evolved over time: students have taken ownership of the game and filled in its gaps in creative ways. Indeed, the lack of explanation on the concepts in the game made some students feel uncomfortable because they did not master them enough, so players got used to seeking the information in their course. They search in the online version of the course, which they access either on their cell phone or on their computer. Most of teams switch between the physical game and the computer screen at every turn of the game. Even though we didn't design the game with this idea in mind, the students made it a *de facto* hybrid version (cf. figure 6).

In addition, as mentioned above, feedback from and discussions with students highlighted the need to show more parts of the software interfaces and to allow to better understand how their interactions work.

As a solution to this need, we imagined an enhanced version of JADE (cf. figure 12): the software boards of JADE V3 include a QR code that students can scan to, depending on the software, watch a video that demonstrates interactions with the software or test the software directly themselves. This hybrid version of JADE meets the identified need while remaining simple enough to implement by the game designer and to use by the players. Moreover, the digital part of the game is optional, so only the players who find it necessary and who appreciate the used technology utilize it. Indeed, we want JADE to be able to remain an unplugged game for our over-connected students.

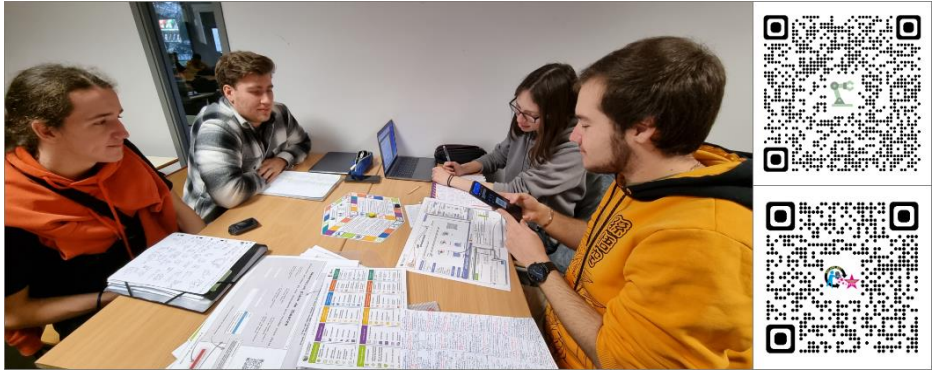

**Fig. 12.** First use of hybrid JADE: Alexandre is scanning a QR code to find out more about how "SOS-info" works (left) and examples of QR codes (right).

As a first step, in order not to remake all the sets, we created and printed QR codes corresponding to the various software boards to be enhanced and stuck them on all the concerned boards. In a second step, we added by design QR codes in all new software boards created.

The enhanced version of JADE is used since the autumn semester of 2022. The feedback from these first tests is very positive: the enhanced version of JADE fully meets the need, and the students appreciate the device, well beyond our expectations. Indeed, after the intensive use of QR codes in France around the health crisis of Covid-19 (declaration of vaccination, test certificate, presence in a frequented place, etc.), we were afraid that the students would be bored with this device, but it was not the case: against all expectations, they found them very funny.

We observed that the transitions from digital devices to the physical game were very fluid, with students moving from one to the other as they needed. For example, all players work on the software boards together, then one team looks for an additional detail on the computer version or tests the mobile usability on a cell phone, then, once they have found the desired information, they show it to the other players on the device, or if possible, directly on the software board. In our most recent use of JADE V3, it seems that the hybrid version has increased the interactions between players and with the contents.

## 5 Use of JADE for Training

**Contexts of Use.** Since 2018, the game has been used in several contexts: mainly in computer science degree courses, but also in a few bioinformatics master's courses and finally during intensive training sessions on ergonomics for professional computer engineers. Table 1 lists the uses that were made of the different versions of JADE by indicating the period concerned, the type of learners among the three contexts, the version of JADE used, the number of learners and the duration of the session (for durations exceeding 2 hours, the table indicates the sum of the durations of the different sessions with the same students) and finally the cumulative time (in student-hours). In total, the different versions of JADE were used by 868 learners, 725 of whom were different (i.e., removing learners who participated in several different sessions) for 51 hours, for a cumulative total of 2479 student-hours.

**Table 1.** List of uses of the different versions of JADE for training in various contexts.

| period | type of learners | version of JADE | number of students | duration of session | cumulative time |
|---|---|---|---|---|---|
| fall 2018 | computer science degree | V1.0 (2 sets) | 73 | 1,5 | 109,5 |
| spring 2019 - group 1 | computer science degree | V1.1 (4 sets) | 67 | 1,5 | 100,5 |
| spring 2019 - group 2 | computer science degree | V1.2 (4 sets) | 39 | 1,5 | 58,5 |
| mai 2019 | IT engineers | V1.2 | 24 | 3,0 | 72,0 |
| july 2019 | computer science degree | V1.2 | 6 | 1,5 | 9,0 |
| november 2019 | IT engineers | V1.2 | 9 | 2,0 | 18,0 |
| fall 2019 | computer science degree | V2.0 (4 sets) | 17 | 1,5 | 25,5 |
| fall 2019 | computer science degree | V2.1 (8 sets) | 69 | 1,5 | 103,5 |
| spring 2020 | computer science degree | V2.2 (18 sets) | 58 | 1,5 | 87,0 |
| fall 2020 | computer science degree | V2.2 | 68 | 1,5 | 102,0 |
| fall 2021 | computer science degree | V2.2 | 64 | 6,0 | 384,0 |
| winter 2021 | master in bioinformatics | V2.2 | 23 | 2,0 | 46,0 |
| spring 2022 | computer science degree | V2.2 | 68 | 5,5 | 374,0 |
| fall 2022 | computer science degree | V2.2 | 78 | 4,5 | 351,0 |
| october 2022 | IT engineers | V2.2 | 10 | 6,0 | 60,0 |
| winter 2022 | master in bioinformatics | V2.2 | 19 | 2,5 | 47,5 |
| fall 2022 | computer science degree | V3 hybrid (18 sets) | 35 | 1,0 | 35,0 |
| winter 2023 | master in science mediation | V3 hybrid | 18 | 1,5 | 27,0 |
| spring 2023 | computer science degree | V3 hybrid | 99 | 4,5 | 445,5 |
| spring 2023 | computer science degree | V3 hybrid | 24 | 1,0 | 24,0 |
| spring 2023 | master in science mediation | V3 hybrid | 12 | 1,5 | 18,0 |
| june 2023 | IT engineers | V3 hybrid | 11 | 3,0 | 33,0 |
| spring 2023 | computer science degree | V3 hybrid | 6 | 1,0 | 6,0 |
| total | | | 868 (725) | 51,5 hrs | 2479,5 hrs |

**Types of Use.** The main use of JADE is in practical sessions throughout the course, but the game is also used in support sessions or in preparation for the exam. Since version 2.2 of JADE, we use the game four times with the same students during the semester: lime green JADE during the first ergonomics practical session, purple JADE during the third session, emerald JADE during the fifth sessions, and then for the last practical session, which takes place just before the exam. This is the configuration that we found most suitable for our students. There are a sufficient number of sessions dedicated to

JADE so that the students are used to the game and benefit from the advantages of game-based courses, but not too many sessions to leave space for traditional practical sessions and not to tire students.

**Interactions During Sessions.** The outcomes of the use of JADE are very positive on the interactions during the sessions. Indeed, the students interact obviously much more than during a theoretical course, but also much more than during a classic practical session, even if carried out in pairs. The game situation forces the students to remain active during the session and provokes numerous exchanges around the targeted knowledge: between peers and with the teacher/game master. Moreover, the students receive immediate feedback throughout the game: from their teammate, from the players of the opposing teams and from the teacher.

**Debriefing.** As recommended by research on the use of games in educational contexts, the session with JADE ends by a debriefing addressing the five dimensions proposed by Plumettaz-Sieber et al. [31]: the affective aspects of their experience with the game, the awareness of the knowledge addressed making the pedagogical objectives explicit, the decontextualization outside the game, the legitimization of the knowledge handled and its generalization. This is done at the end of the session, either by the teacher orally or *via* the "what next?" section of the "JADE recipe" sheet.

**Contribution to Students' Skills.** Sessions with JADE allow students to address a large range of skills, both directly related to the course and more general: learning/consolidating and applying in practice all the ergonomics concepts taught in the course and arguing one's answer. We do not have quantitative measurements, but over the last few years we have observed that students' engagement was greater, that they learn ergonomics concepts involved more quickly and in greater numbers with JADE than during a traditional practical session. The interaction between students seems to pull up the weaker students and helps to improve the knowledge of the latter.

**Contribution to Students' Well-being.** JADE adopts an innovative playful approach: indeed, proposing a physical board game at the university is a break with the current technology race. This approach is all the more unusual for a computer course. Moreover, JADE insists on the return to face-to-face activities to (re)create links between and with students. Finally, JADE allows students to put their knowledge into practice in a different, unusual, more relaxed, funny, and particularly supportive context. The game is also an opportunity to have a different relationship with teachers, to see them differently.

## 5 Discussion

**Level of Hybridization.** In our opinion, the level of hybridization of JADE V3 is rather low, as there are no scripted interactive activities in the digital part of the game, but on the one hand some simple videos are available (on which the players can however pause, move forward/backward, or adjust the speed) and on the other hand players may freely interact with web applications. However, we think this is the right level of hybridization for the game and its players. Indeed, we think it is important not to use technology for its own sake, but in response to a need, which we believe we have done in this project.

**Reference to the Courses.** We had initially planned to add to the game cards representing the different concepts (with their name, the corresponding family of concepts, a definition, an example of good application and an example of bad application). Due to the time required to create them, we have not implemented this idea for the moment. But progressively with the use of JADE in session, the uses settled, and we observed that the students set up a complementary device to the game, in an instrumentalization process of Rabardel's instrumental genesis [36]: they displayed the corresponding courses on their cell phones or laptops. The best configuration they imagined and set up is the one shown on the left-hand picture in Figure 6: a laptop at the back of the table showing to all players of the team the course slide describing the concept being discussed. This device seems very well suited to the game, providing a flexibility that the creation and printing of cards does not allow. We thus decided to keep this other facet of hybridization between physical game and digital information. However, we could go further and make this hybridization more explicit by giving players tools to make the connection more easily between the game concepts and the course materials. For the moment, the link is only made by the color codes of the concept families and the pictograms representing the concepts.

**Scope of Application.** We believe that the JADE game design is transferable to other domains. The framework for applying the JAD... (acronym for Jeu d'Apprentissage De, which means "game to learn") approach is:
- a course with a set of concepts that need to be understood, learned, and also need to be practiced;
- concepts that can be more or less well applied/respected, for which there is not a unique solution to a problem;
- concepts for which there are arguments to justify the right application/respect, and for which different points of view can coexist;
- concepts that can be materialized/located on a printable illustration.

**JAD'edu.** We have thus already successfully applied the game principle to another field: training students to identify and argue about the different characteristics and functionalities of educational software, as part of an optional course in computer science and mandatory course for future computer science teachers.

**Limits.** Even if we see a lot of advantages to the JADE game, we are aware of the following limitations. First, we have already discussed the issue of students without a board gaming culture: it may be more difficult to motivate these students to become immersed in the game. Secondly, we have noticed that too much use of JADE can become counterproductive. Indeed, during one semester we replaced almost all the practical sessions by sessions with JADE and we observed a progressive lassitude among the students. We have therefore opted for a limited number of sessions with JADE, alternating a classical session with a game session. Thus, the sessions with the game are more appreciated and more effective. Finally, the good functioning of the sessions with JADE are also dependent on the motivation of the teachers / game masters, so it is important to maintain this motivation too.

## 6 Conclusion

As an answer to our twofold problematic "How can we both facilitate the learning of ergonomics and increase the motivation of students" and "How can we adopt a playful approach at university while encouraging direct interpersonal interactions", we proposed a pedagogical game to encourage learning and the practical application of ergonomics concepts.

This game is a physical game (research question 1) that relies heavily on human interaction to promote interaction (research question 3): between partners within a same team, with opposing teams and also with teachers. After several tests, we gradually opted for 3 game sessions during the semester, alternating sessions with JADE and classic practical sessions (research question 2).

In this paper, we presented JADE, our educational board game which allows learners to practice software ergonomics. We also presented the hybrid version of JADE which enriches the initial game by exploiting QR codes coupled with videos or interactive software. Since 2018, the game has been used by almost 850 different students for more than 2500 hours in cumulative time. At a time when digital serious games are everywhere, in a course where everything is computer based and dematerialized, the return to physical objects and human relations in an unplugged social activity is unusual but appears to be highly effective and much appreciated. This activity is complementary to the other pedagogical activities of the course and is very beneficial to both students and teachers.

**Benefits.** The feedback both from the students and from teachers is very positive. Indeed, the impact of the game on students' motivation, engagement and understanding of the taught concepts and their relations is very good. As well we find that the use of games can have an impact on teacher motivation and engagement. We also believe that JADE has an impact, beyond our course, on the students' overall motivation, and also on group cohesion and sense of belonging to a class. Thus, JADE contributes to a good image of the course, a good image of the degree, or even a good image of the university.

**Unforeseen Consequences.** We realized that making the game forced us to structure the course more to allow students to easily make the link between the course and the game. The students also noticed the synthetic aspect of the game: for example, they asked to have a version of the pictogram legend sheet that lists all the concepts of the course, to use it to revise before the exam.

### 5.2 Future issues

**More Ergonomics Concepts.** We had initially planned to deal with other ergonomics concepts in JADE: concepts that we only mention in class due to lack of time (other families of concepts such as those of Joëlle Coutaz [38], Ben Schneiderman [39], Jakob Nielsen [40] or Jean-Pierre Meinadier [41]). We could add them to the multicolored game board *via* boxes of other colors or create a new game board dedicated to these criteria.

**More Hybridization.** As seen in the discussion, we could imagine a way to facilitate the link between the concepts in the physical game and the online version of the course (*via* the legend sheet). We have not found an operational solution yet. But, in order to give a better access to the legend sheet (that some indelicate students could be tempted to steal to revise for the exam), we have already added a QR code on the new version of this sheet that gives access to its online version. Other improvements of the same nature, in the direction of increased hybridization, are being considered.

**Gamification Dice.** For those players who want more fun in the game, we plan to add the optional use of a "gamification dice" in JADE, which adds new rules to the existing ones. Depending on the draw of the dice after the concept has been chosen, the following changes must be taken into account by the playing team for this turn: (1) the team has 1 penalty point; (2) the team plays 2 consecutive times; (3) the team studies the next concept in the list; (4) the team passes its turn; (5) the team adds up the two dice to move forward; (6) the team's (1st) point scores double. This optional dice will make it possible to adapt the level of playfulness of JADE a little more to the students' appetence for games.

**From JADE to ORANJADE.** Although JADE is deliberately a physical game, periods of lockdown due to the Covid-19 health crisis led us to consider a digital version of JADE. Thus, we created the prototype of ORANJADE (an acronym whose English equivalent could be "gOing to Revise between mAtes with the Numerical version of JADE"), a collaborative educational platform for training in ergonomics. Unlike JADE, ORANJADE is designed to be played independently of teaching, outside of class time.

**Acknowledgments.** This work was carried out in collaboration with people from University Claude Bernard Lyon 1 (UCBL): Fabien Duchateau, associate professor, Pierre Yang and Maëlys Daubias, students. It also has received support from UCBL *via* the Innovative Pedagogical Practices 2018 call for projects. The author would like to thank all the undergraduate computer science students who participated in the various tests and uses conducted with JADE, Nora Van Reeth for the design of the logo, Philippe Daubias for proofreading the paper, and the members of her family who patiently participated in the beta-tests, as well as in the production of the games.